\documentclass[aps,prl,twocolumn,showpacs,amsmath,amssymb]{revtex4}
\usepackage{graphicx} 
\usepackage{dcolumn} 
\usepackage{bm} 
\begin{document}


\title{Stochastic Modelling Approach to the Incubation
Time of Prionic Diseases}

\author{A.~S.~Ferreira$^1$, M.~A.~da Silva,}
\email{maasilva@fcfrp.usp.br}
\author{J.~C.~Cressoni$^1$}
\email{cressoni@fis.ufal.br}
\date{\today}
\affiliation{
$^1$Universidade Federal de Alagoas,
Departamento de F\'{\i}sica,
57072-970 Macei\'o (AL), Brazil\\
$^*$Departamento de F\'{\i}sica e Qu\'{\i}mica,
FCFRP, Universidade de S\~ao Paulo,
14040-903 Ribeir\~ao Preto, SP, Brazil
}

\begin{abstract}

Transmissible spongiform encephalopathies like the bovine
spongiform encephalopathy ($\text{BSE}$) and the
Creutzfeldt-Jakob disease ($\text{CJD}$) in humans are
neurodegenerative diseases for which prions are the attributed
pathogenic agents.
A widely accepted theory assumes that prion replication
is due to a direct
interaction between the pathologic ($\text{PrP}^{\text{Sc}}$) form
and the host encoded ($\text{PrP}^\text{C}$) conformation, in a kind of
an autocatalytic process.
Here we show that
the overall features of the incubation time of prion diseases
are readily obtained if the prion reaction is described by a 
simple mean-field model.
An analytical expression for the incubation time distribution then follows by
associating the rate
constant to a stochastic variable {\it log normally} distributed.
The incubation time distribution is then
also shown to be {\it log normal} and
fits the observed $\text{BSE}$ data very well.
The basic ideas of the theoretical model are then 
incorporated in
a cellular automata model.
The computer simulation results yield the correct $\text{BSE}$
incubation time distribution at low densities of the host encoded protein.

\end{abstract}
\maketitle


The so-called prion diseases comprise a group of fatal transmissible
spongiform encephalopathies (TSE) like the well known 
bovine spongiform encephalopathy (BSE) and sheep scrapie.
In humans, these progressive neurodegenerative diseases include
Kuru, Creutzfeldt-Jakob disease (CJD),
Gerst\-mann-Straeussler-Scheinker syndrome (GSS) and
fatal familial insomnia (FFI).
Common pathology includes spongiform degeneration
and characteristic formation of plaques in the brain tissue~\cite{weissman}.
Variant CJD correlated with a (BSE)-like prion strain,
have been identified and are believed to be linked to the consumption
of contaminated food~\cite{collinge,will,bruce,hill}.

The protein-only hypothesis~\cite{griffith} states that
the infectious agent is a protein, named prion~\cite{prus1,prus2},
which is devoid of nucleic acid and
capable of replicating itself in
the absence of these traditional genetic material.
Two conformations of this protein are important for characterizing
the disease, namely, the normally folded host-encoded cellular
protein called $\text{PrP}^\text{C}$ and an abnormal pathogenic conformation
named $\text{PrP}^\text{Sc}$.
The latter form is hydrophobic, has a
tendency to form aggregates and may be found in different strains.
The pathogenic form $\text{PrP}^\text{Sc}$ is more stable than
the endogenous cellular form, and is known to be resistant
to enzymatic digestion, radiation and high temperatures.
One of the most accepted models for prion replication assumes
that this form acts as a template for converting the host
prion
into its own conformation
in a kind of an autocatalytic reaction~\cite{telling,prus3}.
Understanding the dynamics of the
$\text{PrP}^\text{C} \rightarrow \text{PrP}^\text{Sc}$
transformation is crucial
if one is attempting to explain and predict
the main stages of the disease.
The reaction is complex,
perhaps involving other participants
possibly acting as chaperone, to help mediate protein
folding~\cite{telling2}.
The number of parameters involved for thoroughly describing
the transformation process is thus expected to be very
large~\cite{laurent,eigen}.
It is therefore important to be able to recognize which 
ones are mandatory, i.e., responsible for the major aspects of
the dynamics.

Here we present a simple, analytically solvable, mean-field model
for describing the prion reaction problem,
which focuses on
realistically reproducing the incubation time of the disease.
For notational convenience it is useful to introduce the following
definitions: $A$ stands for the host protein ($\text{PrP}^\text{C}$)
and $B$ stands for the pathogenic form ($\text{PrP}^{Sc}$)
with $a=[A]$ and $b=[B]$ denoting volume concentrations.
We then write the autocatalytic conversion reaction simply as
\begin{equation}
A + B ~~~\stackrel{K}{\longrightarrow} ~~~2B \label{eq:rea}
\end{equation}
where $K$ is the reaction rate.
For simplicity we shall assume that the total concentration $a+b=\rho$
is kept fixed at all times.
This means that there is no metabolic decomposition of
$B$ and any metabolic decomposition of $A$ is
immediately compensated by the host genetic system. It also implies
that the host takes no action for producing new, normal
protein, as the reaction takes place.
In order to stick to the simplest possible case
we are also assuming
that the reaction is unidirectional and
favors the most stable form $\text{PrP}^\text{Sc}.$
No other strains are supposed to be present
and both forms are assumed to be uniformly distributed. 
The kinetic evolution~\cite{murray}
is then given by \mbox{$db/dt=K a b=K(\rho-b) b$}
which is the simplest possible nonlinear equation
describing an autocatalytic reaction.
This equation can be easily integrated up to the time $T$
giving
\begin{equation}
T = \frac{1}{K(a_0+b_0)}
\ln \left[\frac{a_0}{b_0}\left(\frac{b(T)}{a_0+b_0-b(T)}\right )
\right] \label{eq:tI}
\end{equation}
with $b_0$ being the infection dose given at time $t=0$ and
$a_0$ the initial concentration of $A$.
According to this expression $b(t)$ is slowly varying for small $t$,
followed by a period of rapid increase in a short time interval,
then reaching a plateau for long enough times when the reaction
stops~\cite{nos,laurent}.

\begin{figure}
\includegraphics{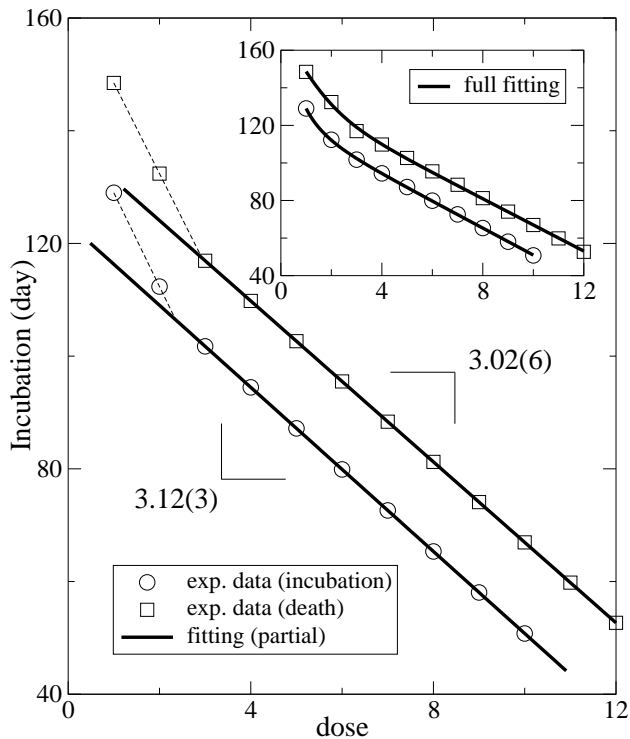}
\caption{\label{fig1}
Dependence of the incubation time ($T_I$) on the infection
(initial) $\mathit{dose}$ ($=n$ with $b_0/\beta_0=10^{n-10}$).
The experimental data
were obtained from Prusiner's work~\cite{prus4}
and dashed lines are just meant to lead the eye.
In the main figure we apply regression to the data ($n>2$) to obtain the
best fit with eqn.~(\ref{eq:tI2}).
The most diluted part ($n\lesssim 2$)
was left out due to the abrupt change in behavior in this region,
leaving only two points ($n=1,2$) for the fitting.
Therefore only the inclinations ($=1/(Ka_0)$) are kept.
The inset shows a non-linear least-squares full fitting (all $n$)
to eqn.~(\ref{eq:tI2}) with the ansatz 
$K \rightarrow K_{\mathit{eff}}=K[1-a_1/(a_2+\mathrm{exp}(n))]$.
}
\end{figure}

We now define
the incubation time ($T_I$) as the time it takes for the
number of pathogenic prions to reach a given value $b_I$, i.e.,
$b(T_I)=b_I.$
(It makes no difference to our calculations whether
$b_I$ represents a number of prions or an aggregate with
size $b_I.$)
A useful approximation can be obtained
by assuming, reasonably, that
\mbox{$b_0/a_0 << 1$}. This gives

\begin{equation}
T_I \simeq \frac{1}{K a_0}
\ln \left[\frac{b_I}{b_0}\left(\frac{1}{1+b_I/a_0}\right )
\right] ~~~~~~. \label{eq:tI2}
\end{equation}
This log-dependence of the
incubation time on the initial dose
was quantitatively
observed by Prusiner~\cite{prus4}
from inoculation of a form of scrapie in hamsters (Fig.~\ref{fig1}).
Prusiner's results
also indicate that the survival time is practically independent
of the dose.
Eqn.~(\ref{eq:tI2}) is consistent with
this finding~(see also \cite{nos}).
If we define the time of death 
as the time it takes for the number of $B's$ to reach
the value $b_D$, i.e., $b(T_D)=b_D$, we find
that $T_D-T_I$ does
not depend on $b_0$. Moreover, eqn.~($\ref{eq:tI2}$)
can be easily adapted to fit Prusiner's data.
In order to mimic the \mbox{end-point} titration method used in
the experiment, we first define all concentrations relative to
the largest experimental concentration which we shall call $\beta_0.$
We then write $b_0/\beta_0=10^{n-10}$ ($n$ = $\mathit{dose}$)
and allow $n$ to vary from $n=0$ (smallest concentration) to
$n=10$ (largest concentration).
We can now apply regression to the data (using only the
integral values for $n$) to obtain the best fit.
Notice, however, that the experimental curves are composed of two
branches, both exhibiting a
sudden increase in the inclination  for $n\lesssim 2$
(see Fig.~\ref{fig1}).
This behavior seems to be indicative of a threshold,
possibly leading to a smaller rate constant at high dilutions.
One can simulate a
{\it dose} dependent activation mechanism linked to the rate
constant $K$ by making the following ``ansatz'':
we make $K \rightarrow K_\mathit{eff}$
with $K_\mathit{eff}=K[1-a_1/(a_2+\exp(n))]$.
With these implementations, eqn.~(\ref{eq:tI2})
reads \mbox{$T_I=C-[\ln10/(Ka_0)]n$}, $C$ being
a constant (independent of $b_0$).
The phenomenological constants, estimated with
a non-linear least-squares fitting to this equation,
with $K$ replaced by $K\mathit{eff}$, are
$a_1=0.23(4)$ ($0.61(2)$) and $a_2=-0.51(6)$ ($2.1(2)$)
for the incubation (death) curve.
The result of the full fitting is shown in the inset of Fig.~\ref{fig1}.
Notice that $K_\mathit{eff}$ rapidly approaches $K$ for $n>2.$

However, we decided to avoid dealing with the controversial features
associated with the region $n\lesssim 2$ (containing only two
points) and stick to the (larger) less inclined part of the
experimental curve. Therefore
any parameter obtained from the
\mbox{$y-$intercepts} in Fig.~\ref{fig1} will not be
taken into account.
The regression coefficient gives
$1/(Ka_0)=3.12(3)$~days for the
incubation part of the curve and $1/(Ka_0)=3.02(6)$~days
for the death part of the curve. This (partial) fitting is
represented by the continuous line in the main part of Fig.~\ref{fig1}.
We can easily check the reasonableness of these figures.
Notice that we could have started with the
Michaelis-Menten equation, namely,
\mbox{$db/dt=K_T[ab/(K_M +a)]$} with $K_T$ and $K_M$ being the turnover
number and the Michaelis constant respectively~\cite{eigen}.
Direct integration of this equation yields

\begin{eqnarray}
K_T\times T &=& \frac{1}{a_0+b_0}
\left[
K_M \ln \left(\frac{a_0}{a_0+b_0-b(T)}\right) \right. \nonumber \\
&+&
\left. (K_M + a_0 +b_0)\ln \frac{b(T)}{b_0}
\right]
\label{eq:mm}
\end{eqnarray}
which is consistent with eqn.~(\ref{eq:tI})
for $K_M >> a_0 >> b_0$ and $K\simeq K_T/K_M.$
We can therefore estimate the prion $K_T/K_M$ ratio
for the scrapie strain used by Prusiner in hamsters.
If we assume $a_0\sim$~nanomole~liter$^{-1}$~\cite{laurent,eigen}
we find $K_T/K_M \sim 10^3~\mathrm{M}^{-1}\mathrm{s}^{-1}.$
This value is within the range expected for enzymes, in which case
$K_M$ lies between 10$^{-7}$~M to 10$^{-1}$~M
and $K_T$ falls in the range from 10~s$^{-1}$ and
10$^7$~s$^{-1}.$

Having discussed the behavior of the incubation time on $b_0$,
we now turn our attention to the dependence of $T_I$ on $a_0$.
The role played by the host prion initial
concentration is useful for
describing reactions, such as (\ref{eq:rea}), in
numerical simulation approaches.
The explicit power law dependence of $T_I$ on $a_0$ can
be seen by expanding eqn.~(\ref{eq:tI}) in terms
of $b_I/a_0.$ This gives
\begin{equation}
T_I  \sim \frac{1}{Ka_0}
\left\{
\ln {\frac{b_I}{b_0}} + \frac{b_I}{a_0} +
{\mathcal O}\left( ({\frac{b_I}{a_0}})^2 \right)
\right\}
\label{eq:tI_app}
\end{equation}
\noindent
and therefore $T_I\sim A_1/a_0 + A_2/a_0^2.$
This kind of behavior,
having the form of a sum of monomer and dimer terms,
has already been suggested in the literature~\cite{cox}.
However, the determination of the explicit dependence of the
coefficients $A_1$ and $A_2$ on $b_I$ and $b_0$, as shown here,
was only possible because of the simplicity of
the model.

\begin{figure}
\includegraphics{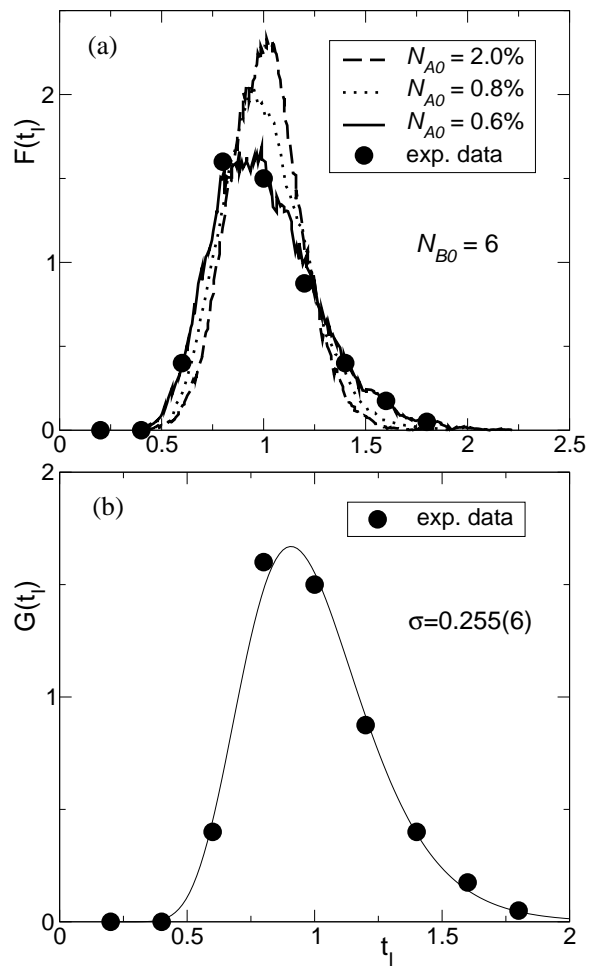}
\caption{\label{fig2}
Incubation time distributions with the time scale
normalized by the mean incubation time, i.e., $t_I=T_I/\overline{T}_I.$
The full circles represent the observed incubation time distribution
for BSE-infected cattle in UK~\cite{cox,stekel,anderson}.
Fig.~\ref{fig2}(a) shows the results from computer
simulations on an $N\times N$ (${N}=200$) square lattice,
with a number of $\text{PrP}^\text{Sc}$ seeds
$N_{B0}=6$ (see text).
The values of $N_{A0}$ shown represent the initial
$\text{PrP}^\text{C}$ concentration and only a few curves for
$N_{A0}$ were drawn to avoid figure cluttering.
Notice the tendency for better agreement with
the observed results as $N_{A0}$ gets smaller.
Fig.~\ref{fig2}(b) shows the same experimental data
as in $(a)$, along with the proposed analytical distribution
$G(t_I)$,
obtained from our model assuming a {\it log normal} distribution for
the rate constant.
When the time units are scaled by the mean time we are left
with a single parameter, namely $\sigma$, whose best
fitted value is given by $\sigma=0.255 \pm 6.$
}
\end{figure}

The initial concentration of the endogenous $\text{PrP}$ protein is
determinant for the dynamics of the prion reaction since it
represents the reaction fuel. The higher the initial concentration
$a_0$, the lower the time for the $\text{PrP}^\text{Sc}$ concentration to
reach the value $b_I.$
These results have been obtained through
careful computer simulations by Cox {\it et al.}~\cite{cox}.
They also showed
that the incubation time distributions for different $a_0$
collapse to a single form
if the time scale is properly normalized to unity.
Will our simple, minimally parametrized model
represented by the basic reaction (\ref{eq:rea}),
be able to reproduce such results?
In order to address this question,
we ran computer simulations based on a
cellular automata ($\mathit{CA}$)
with rules following
a close resemblance to our
model.

According to the {\it CA} rules,
an $N\times N$ ($N=200$) square lattice is
randomly populated with a number ($N_{A0}$) of the host
$A=\text{PrP}^\text{C}$ protein
and a number ($N_{B0}=6$) of the $B=\text{PrP}^\text{Sc}$
misfolded protein.
$N_{A0}$ is given as a small percentage of the total number
of sites available and to each of the $B$ sites
is assigned a ``mass'' ($m$), initially set to unity.
The $A's$ and $B's$ are allowed to diffuse 
randomly to their nearest neighbor sites and a reaction occurs when
a $B$ is approached by an $A$
at a distance $d\leq \sqrt {m}.$
In this case the normal Prion disappears and the misfolded Prion has its
mass increased by one. The reaction is unidirectional, favoring $B$,
with the $A's$ slowly disappearing from the system,
keeping $A+B=\mathit{constant}.$
One site-by-site sweep through the lattice is made
for diffusion followed by another one for reaction. The time 
unit is then increased by one (arbitrary units).
The reaction stops when one of the
masses reaches the value $m=40$, the corresponding computer time thus
characterizing the incubation time.
The above values for the parameters (not the {\it CA} rules)
were adjusted from the numerical simulations of
Cox~{\it et al.}~\cite{cox} in a hexagonal lattice.
The mass is here to mimic clusterization
without assigning any geometric form to the cluster.
Besides simplifying the computer code and speeding up the
simulations, this helps reducing the influence of local
topology on the final results.

Fig.~\ref{fig2}(a) shows the simulation results for
the incubation time distributions for
several values of $N_{A0}$,
with the time scale normalized by the mean time.
Notice that as the $\text{PrP}^\text{C}$
concentration is decreased,
the corresponding distribution converge asymptotically 
to the experimental results
(BSE-infected cattle in UK~\cite{cox,stekel,anderson})
represented by the full circles.
Increasing $N_{A0}$ makes the system more homogeneous which
diminishes fluctuations and narrows the distribution.
The biological concentrations (believed to be nanomolar)
correspond to an areal concentration around
$N_{A0}^{bio}=0.001\%$~\cite{cox}.
With the $\mathit{CA}$ rules above, such
small concentrations would require very large computing
time, if feasible at all.
The best agreement is obtained for $N_{A0}=0.6\%$ which is
far as we could go with these simulations.

Our next issue is to search for an
analytical form for the incubation time distribution.
Knowledge of such a function is not only
important to check the reliability of the model but also
to provide a distribution that can be used in statistical
studies~\cite{anderson2}.
We need to adapt the deterministic model to accommodate
a stochastic variable
following a known distribution and associate it with
eqn~(\ref{eq:tI}) (or (\ref{eq:tI_app})).
Since the protein-folding process actually involve many
steps~\cite{eigen}, possibly chaperone-assisted~\cite{liautard},
the end result of the prionic reaction
can adequately be viewed as
a series of multiplicative processes.
It is therefore reasonable to assume that
the distribution of the reaction rate $K$ in a population
is {\it log normal}~\cite{shlesinger}.
Since $K\propto 1/T_I$ 
it is easy to show that
$T_I$ also follows a {\it log normal} distribution with the same deviation.
The scaled distribution $G(t_I)$, 
with $t_I = T_I/\overline{T}_I$, is then readily obtained. One finds
\begin{equation}
G(t_I)=\frac{1}{\sigma \sqrt{2\pi}} t_I^{-1}
\exp\left[-\frac{1}{2}\left(
\frac{\ln t_I +(5/2)\sigma^2}{\sigma}
\right)^2
\right]
\label{eq:d2}
\end{equation}
which does not depend neither on the initial variables nor on $b_I.$
We are therefore left with a single
fitting parameter, namely $\sigma$, the standard deviation
of $\ln K.$
Applying non-linear least squares fitting to Eqn.~(\ref{eq:d2})
we get $\sigma = 0.255 (\pm 6).$ The final result is
shown in Fig.~\ref{fig2}(b).
In this Figure
the observed data are the same used in reference~\cite{cox} for
BSE-infected cattle in the United Kingdom born in
1987~\cite{stekel,anderson}.

In conclusion, a simple mean field model, based on an
autocatalytic mechanism, is shown
to contain the basic ingredients necessary to describe
the essential features associated with the incubation time of
the complex prion conversion reactions.
Assuming that the rate constant is a random variable,
following a {\it log normal} distribution, we were able to
provide a closed form for the incubation time distribution
of BSE-infected cattle.
The surprisingly simple analytical expression derived
for the incubation time distribution contains only one
parameter, namely the variance of the logarithm of the
rate constant.
The simplicity of the model is characterized by
the almost naive differential equation upon which it is based,
by simple computer simulations, and by the minimal
set of parameters used to describe the autocatalytic process.

J.C.C. and A.S.F acknowledge funding from CNPq (476376/2001-7).
J.C.C. is very grateful to Profs. R.~J.~V. dos Santos
and S.~B. Cavalcanti for discussions. We
particularly acknowledge Prof. M. L. Lyra for
fruitful suggestions and careful reading of the
manuscript.

\cleardoublepage 
\end{document}